\documentclass[11pt]{article}

\usepackage[T2A]{fontenc}
\usepackage[english]{babel}

\usepackage{epigraph}
\usepackage{indentfirst}

\usepackage[unicode=true]{hyperref}
\usepackage{xr}
\usepackage{latexsym}
\usepackage{amsfonts}
\usepackage{amsmath}

\font\twmsbm=msbm10 scaled 1200 \font\nmsbm=msbm9
\newfam\bbold
\textfont\bbold=\twmsbm \scriptfont\bbold=\nmsbm
\scriptscriptfont\bbold=\nmsbm

\font\twscr=rsfs10 scaled 1200 \font\nscr=rsfs10
\newfam\script
\textfont\script=\twscr \scriptfont\script=\nscr
\scriptscriptfont\script=\nscr

\newcommand{\be}{\begin{equation}}
\newcommand{\ee}{\end{equation}}
\newcommand{\bean}{\begin{eqnarray*}}
\newcommand{\eean}{\end{eqnarray*}}

\newcommand{\bea}{\begin{eqnarray}}
\newcommand{\eea}{\end{eqnarray}}

\newcommand{\diag}{{\rm diag}}

\newcommand{\p}{\partial}

\author{Mikhail G. Ivanov
\\ $~~~$\\
\small{Moscow Institute of Physics and Technology (state university)}\\
\small{Institutski per. 9, 141700, Dolgoprudny, Moscow region,
Russia }\\ \tt{e-mail: mgi@mi.ras.ru} }

\title{Superluminal motion and Lorentzian symmetry breaking
 and repairing in two-metric theories}
\date{February 18, 2012}

\begin{document}
\maketitle

\begin{abstract}
The new results by OPERA collaboration claim the discovery of
superluminal neutrinos. Superluminal particles have to break
Lorentzian symmetry or causality principle. The method discussed
gives us the possibility to reintroduce Lorentzian symmetry without
breaking of causality.
\end{abstract}

\section{Lorentzian symmetry or causality?}
New measurements of high energy neutrino speed by OPERA
collaboration \cite{OPERA} suggest the superluminal motion, i.e.
motion of particle along space-like world line. Due to theoretical
difficulties the discussion of superluminal neutrinos generates the
great number of criticism and hypotheses (see \cite{IVV} and
references therein).

If the Lorentzian symmetry still valid, all space-like directions of
space-time are equivalent. The possibility of superluminal motion
means the possibility of motion backward in time. Actually \emph{it
is backward time motion} in some high-speed inertial systems.

So, if one believes in superluminal motion, one has to choose what
principle he eliminates: \emph{Lorentzian symmetry or causality}.
The causality looks more important. Nevertheless, Lorentzian symmetry
is extremely useful tool, which lays in basis of modern physics.

\section{``Lorentzian'' symmetry of elastic media and its breaking}

Let us imagine perfect elastic media with sound waves described by
wave equation
\be \label{sound}
  \left(\frac{\p^2~}{\p t^2}-u^2\triangle\right)\mathbf{w}=0,
\ee
where $u$ is speed of sound, $0<u<c$.

Sound waves equation has the sonic ``Lorentzian'' symmetry with
speed of sound instead of speed of light. Sonic ``Lorentzian''
symmetry preserve the ``sonic interval'' (``sonic metric'')
$$
ds_{sonic}^2=-u^2dt^2+dx^2+dy^2+dz^2=h_{MN}\,dX^M\,dX^N.
$$
This sonic ``Lorentzian'' symmetry is broken by non-elastic
phenomena (e.g. light). Non-elastic phenomena make possible
supersonic motion.

The true Lorentzian symmetry preserve the ``true interval'' (``true
metric'')
$$
ds_{true}^2=-c^2dt^2+dx^2+dy^2+dz^2=g_{MN}\,dX^M\,dX^N.
$$
Existence of the elastic medium breaks the true Lorentzian symmetry
of space time.
 It provides us with preferable coordinate system, the
system of medium.

Actually the medium breaks true Lorentzian symmetry at \emph{the
level of solutions} of field equations. At \emph{the level of field
equations itself} the true Lorentzian symmetry is still valid
(``repaired'').

To repair Lorentzian symmetry,
let us introduce 4-velocity of elastic medium $U^M$ ($U^MU_M=-1$,
$U_M=U^Ng_{NM}$). Projectors $P_{MN}$ (to vectors parallel to $U^M$)
and $\bar P_{MN}$ (to vectors orthogonal to $U^M$) are defined
$$
  P_{MN}=-U_MU_N,\qquad \bar P_{MN}=g_{MN}-P_{MN}.
$$
Sonic metric $h_{MN}$ is defined in terms of true metric $g_{MN}$ and 4-velocity
through the projectors
$$
  h_{MN}=\frac{u^2}{c^2}\,P_{MN}+\bar P_{MN},\quad
  h_{MN}h^{NK}=\delta_M^K.
$$
The covariant (Lorentzian symmetric) form of equation \eqref{sound} is
\be\label{sound2}
  \nabla_M(h^{MN}\nabla_N)\mathbf{w}=0.
\ee

\section{``Elastic medium'' of space-time}

Similarly to breaking of sonic ``Lorentzian'' symmetry and
preserving of ``true'' Lorentzian symmetry in the elastic media one
could think on breaking Lorentzian symmetry by superluminal
neutrinos.

It requires introducing of some ``medium'', which generates ``light
metric'' $h_{MN}$, like ``sonic metric'' in the example above. This
light metric is the metric, which interact with almost all sorts of
matter, so it dictates the maximal speed of almost all interactions,
i.e. the speed of light $c$. The light ``Lorentzian'' symmetry
preserves the light metric.

Light Lorentzian symmetry could be broken by some sorts of matter,
including neutrinos. Nevertheless, there is also the ``true
Lorentzian'' symmetry, which preserves the ``true metric'' $g_{MN}$.
The true metric dictates the other maximal speed $c'>c$.

One could rewrite all the standard Lagrangians using light metric
for luminal and subluminal matter and true metric for superluminal
matter.

The ``medium'', which generates light metric plays the role
partially similar to historically banned luminiferous aether.

One could easily construct large number of two-metric models
starting from different models of relativistic elastic media. The
geometrical method of relativistic elastic media modeling is
developed in the series of papers \cite{MGI1}--\cite{MGI9}.

Different two-metric approaches to interpretation of OPERA results
were considered in papers \cite{1109.6005}, \cite{1109.6298},
\cite{1109.6930}, \cite{1110.0697}, \cite{1110.1330}.

\section{The example}
   Let us sketch the simplest two-metric model.

   To describe ``medium'' we use one scalar field $\tau$
 (``luminiferous scalar'') with light-like gradient.

   The light metric $h_{MN}$ specified by
 relation\footnote{E.g. if
$g_{MN}=\diag(-c^{\prime2},1,1,1),\quad \tau=v\cdot t,\quad v<c',$\\
 then
$h_{MN}=\diag(-c^{\prime2}+v^2,1,1,1)
  =\diag(-c^2,1,1,1),\quad c=\sqrt{c^{\prime2}-v^2}.
$}
\be\label{h_mn}
  h_{MN}=g_{MN}+(\p_M\tau)\,(\p_N\tau).
\ee
  The inverse light metric is
\be\label{h-inv}
  h^{MN}=g^{MN}-\frac{(\p^M\tau)(\p^N\tau)}{1+(\p_K\tau)(\p^K\tau)},
\ee
 where
$$
  g_{MN}g^{NL}=\delta_M^L,\qquad
  h_{MN}h^{NL}=\delta_M^L,\qquad
  \p^M\tau=g^{MN}\p_N\tau.
$$

  The ``regular'' fields $\varphi$ are described by standard field theory
  action $S_{standard}[\varphi,g_{MN}]$ with true metric $g_{MN}$ replaced by
  light metric $h_{MN}$
$$
  S_{regular}[\varphi,g_{MN},\tau]
  =S_{standard}[\varphi,g_{MN}\to h_{MN}].
$$
  We have the following relation for the variation of light metric
$$
  \delta h_{MN}=\delta g_{MN}
  +(\p_M\delta\tau)\,(\p_N\tau)
  +(\p_M\tau)\,(\p_N\delta\tau).
$$
 So, all the ``regular'' fields described by the standard field
 equations with ``true'' metric $g_{MN}$ replaced by ``light'' metric
 $h_{MN}$:
$$
  \frac{\delta S_{regular}}{\delta \varphi}
 =\left.
  \frac{\delta S_{standard}}{\delta \varphi}
  \right|_{g_{MN}\to h_{MN}}.
$$
  The gravitational field also could be considered as ``regular''
 field, \emph{if there is no non-regular fields}
$$
  \frac{\delta S_{regular}}{\delta g_{MN}}
 =\left.
  \frac{\delta S_{standard}}{\delta g_{MN}}
  \right|_{g_{MN}\to h_{MN}}.
$$
  The standard action obeys the Lorentzian symmetry,
so all the regular fields obey ``light'' Lorentzian symmetry.

  To break the light Lorentzian symmetry one could introduce
some ``non-regular'' fields $\phi$, which interacts with true metric
$g_{MN}$.
\be\label{action}
  S[\varphi,g_{MN},\tau,\phi]
 =S_{regular}[\varphi,g_{MN},\tau]
 +S_{non-regular}[g_{MN},\tau,\phi]
\ee

 Field equation for the luminiferous scalar $\tau$ has the form
$$
  \frac{\delta S}{\delta \tau}
 =\p_M\left(2(\p_N\tau)\left.
  \frac{\delta S_{standard}}{\delta g_{MN}}
  \right|_{g_{MN}\to h_{MN}}\right)
 +\frac{\delta S_{non-regular}}{\delta \tau}.
$$
  Due to Einstein equations
$$
  \left.\frac{\delta S_{standard}}{\delta g_{MN}}
  \right|_{g_{MN}\to h_{MN}}
  =\frac{\delta S_{regular}}{\delta g_{MN}}
  =-\frac{\delta S_{non-regular}}{\delta g_{MN}}.
$$

  If we introduce no action for non-regular fields, then ``true'' metric
 and ``luminiferous scalar'' are non-observable, and ``light metric''
 is the only physical metric.

\section{Toy model}
  The simplest two-metric toy model involve one ``regular'' scalar field
$\varphi$ and the only ``non-regular'' scalar $\phi$,
which is small perturbation to luminiferous scalar
\be
\tau=\tau_0+\phi=v\cdot t+\phi.
\ee
True Lorentzian
metric $g_{MN}=\diag(-c^{\prime2},1,1,1)$ is fixed.
Let $(\p\varphi)^2_g=g^{MN}(\p_M\varphi)(\p_N\varphi)$,
$(\p\varphi,\p\tau)_g=g^{MN}(\p_M\varphi)(\p_N\tau)$.
The action \eqref{action} is combination of two scalars
with two metrices \eqref{h_mn},\eqref{h-inv}
\be
  S[\varphi,\tau]
  =\int d^4x\left[-(\p\varphi)^2_h
  -(\p\tau)^2_g\right]
 =\int d^4x\left[-(\p\varphi)^2_g
  +\frac{(\p\varphi,\p\tau)^2_g}{1+(\p\tau)^2_g}
  -(\p\tau)^2_g\right].
\ee
  The velocities of small perturbations of $\tau$ (or $\phi$)
 and $\varphi$ are $c'$ and $c=\sqrt{c^{\prime2}-v^2}<c'$.

 The toy model is probably the simplest one.
 It coincides with the more complex massive vector model
 presented in paper \cite{1110.1330}
 in the case of gradient field, i.e. if $\psi_M=\p_M\tau$.

\section{Conclusions}

The paper demonstrates the method of theory construction, which could
be used for relativistic elastic media models or theories with
superluminal particles. The method admits
correspondence with standard relativistic theories, it preserves
Lorentzian symmetry and causality. The appropriate limit reproduces
the standard field theories with no superluminal motion.
~\\

\textbf{Acknowledgments}

The author is grateful to I.V. Volovich for introducing to the problem and
useful discussion.

The work is partially supported by grants RFFI 11-01-00828-a and NS 7675.2010.1.


\end{document}